\documentclass{article}
\usepackage{amssymb}
\usepackage{amsmath}
\usepackage{graphicx}
\input{tcilatex}
\begin{document}

\title{Emission coordinates for the navigation in space}
\author{Angelo Tartaglia \\
Physics Department (DIFIS) of the Politecnico di Torino, \\
and INFN Torino, Italy \\
e-mail: angelo.tartaglia@polito.it}
\maketitle

\begin{abstract}
A {general approach to the problem of positioning by means of pulsars or
other pulsating sources located at infinity is described. The counting of
the pulses for a set of different sources whose positions in the sky and
periods are assumed to be known, is used to provide null emission, or light,
coordinates for the receiver. The measurement of the proper time intervals
between successive arrivals of the signals from the various sources is used
to give the final localization of the receiver, within an accuracy
controlled by the precision of the onboard clock. The deviation from the
flat case is discussed, separately considering the different possible
causes: local gravitational potential, finiteness of the distance of the
source, proper motion of the source, period decay, proper acceleration due
to non-gravitational forces. Calculations turn out to be simple and the
result is highly positive. The method can also be applied to a constellation
of satellites orbiting the Earth.}
\end{abstract}

\section{Introduction{\ }}

An extremely important problem posed by the needs of navigation through
space and also of movement on the surface of the Earth and close to it, is
the one of positioning with respect to some appropriate global reference
system. Presently this problem is dealt mostly using GPS. Analogous systems
are GLONASS and in the next future the European Galileo system (as well as
others partially deployed or under implementation). All these systems were
initially developed for military purposes, and, in the case of both GPS and
GLONASS, they are still under primary military control. The basic idea
there, is quite traditional: if we are able to measure the times of flight
of electromagnetic signals arriving from a set of at least three sources
occupying known positions, then we can, by means of a triangulation process,
work out the position of the receiver in the same reference frame as the
sources. This approach requires a common global time reference and implies
the emitters to be synchronized with each other and with the user. If the
sources are at least four, you may, in principle, dispense with the clock of
the receiver. In practice a rapidly converging trial and error approach is
used and a clock at the reception point is also needed. As it is also well
known, GPS treats the relativistic aspects of the positioning as
perturbations to a basic Newtonian method \cite{ashby}\cite{sanchez}. If
relativity were neglected the positioning error would turn out to be huge
and unacceptable. The special relativistic effects (influence of the speed
of the satellites and Sagnac effect) and the general relativistic ones (blue
shift of the flying clocks with respect to the ones on the ground) are
separately introduced as corrections. Here we shall expound the essentials
of a positioning system developed from scratch on fully relativistic bases
and using the very space-time as the actual fiducial frame. As we shall see
we describe our approach considering the pulsars as sources, even though the
method is applicable to the case of freely falling clocks in the terrestrial
environment (satellites) also. The idea of using pulsars as clocks has been
considered almost from the year of their discovery. It has been discussed in
various ways because it is appealing, however there are technological
limitations which make things difficult. The point is mainly that the
signals from pulsars are very weak, so that, usually, sophisticated and
large sized devices (radiotelescopes) are needed. To avoid the inconvenience
of the dimensions of the antenna, X-ray pulsars have also been envisaged
\cite{sheikh}: the number of sources is not that big and the signals are
weak too, however the noise is comparatively small also. In our case we
shall consider millisecond pulsars where unfortunately the signal to noise
ratio is really unfavorable. We thought however that comparatively large
antennas could be allowed for space missions. In any case we are not
treating here the technology problems. We have developed a computation
method and a reasonably simple (much simpler than the present GPS method)
algorithm to convert the arrival times of the signals from the pulsars into
practical coordinates of a reference frame at rest with the "fixed" stars.
As we shall see, our method includes the effects of the gravitational field
and allows a spacecraft to know what its absolute motion is with an accuracy
controlled by the precision of the clock it carries, making it independent
from control from the ground.

\section{\protect\bigskip Pulsars}

I am here very shortly reviewing pulsars and their features. Pulsars are
neutron stars generated at the moment of the implosion of the core of a
collapsing star in a supernova event. The pulsar possesses a very high
magnetic field whose axis is in general not aligned with the spin axis of
the star. Matter from an accretion disk, falling towards the surface of the
pulsar, is ionized, then expelled, in the form of charged particles, along
the lines of the magnetic field. The particles flux is accompanied by a
highly collimated beam of electromagnetic radiation. Due to the misalignment
of the magnetic field and the rotation axis of the star, the beam behaves as
the radiation from a lighthouse. When we are in an appropriate position for
seeing it, it reaches us one time per revolution of the star. The neutron
star is spun at very high angular speeds by the conservation of angular
momentum during the collapse. The rotation period of the pulsar coincides
with the return time of the pulses. The period of the known pulsars ranges
from the scale of the millisecond up to several seconds and slowly decays in
time, due to rotational energy loss by the star. The behaviour of the
objects under observation since years is very well known. Even though the
time profile of each single pulse is usually different from the others,
pulsars may be thought of as extremely stable clocks whose accuracy is
overcome only by the best modern atomic clocks. In the long term however
pulsars always carry the day. This is the reason why these objects can be
considered as interesting celestial beacons. The known pulsars so far are
approximately 1800 and are all located in our galaxy (apart from a handful
of X-rays emitting objects in the Magellanic clouds), usually within a few
thousands of light years from the Earth. Many of the stars are in binary
systems, however we shall focus on isolated millisecond pulsars. Isolated
objects may reasonably be considered as "fixed stars" since their proper
motions in the sky show up only on rather long periods of time.

\section{\protect\bigskip Emission coordinates}

In a relativistic framework, space and time are compound together in a
four-dimensional manifold endowed with geometrical properties. The
gravitational field is accounted for as a curvature in the manifold. In
order to localize "events" (points in space-time) four independent numbers
(coordinates) are needed. These may of course be, for example, three
Cartesian coordinates and one time, but any quadruple of independent
parameters would fit as well. Consider now four independent clocks. Suppose,
for simplicity, they are in free fall (geodetic motion) and that they
broadcast their own time (their proper time). An observer who receives the
signals from the four emitters gets, at any moment, the information
concerning the proper emission time of the perceived signal from each
source. The quadruple of these times depends on the space-time position of
the receivers; each quadruple is uniquely associated with each event. This
means that the four times conveyed by the electromagnetic signals are indeed
a good set of coordinates for identifying positions in space and time. It is
also clear now why these coordinates are called \textit{emission}
coordinates. Their use has been analyzed and proposed by various authors
\cite{coll1}\cite{coll2}\cite{coll3}\cite{noi1}\cite{noi2}.

\section{\protect\bigskip Basic frame}

\begin{figure}[tbph]
\begin{center}
\includegraphics[width=10cm]{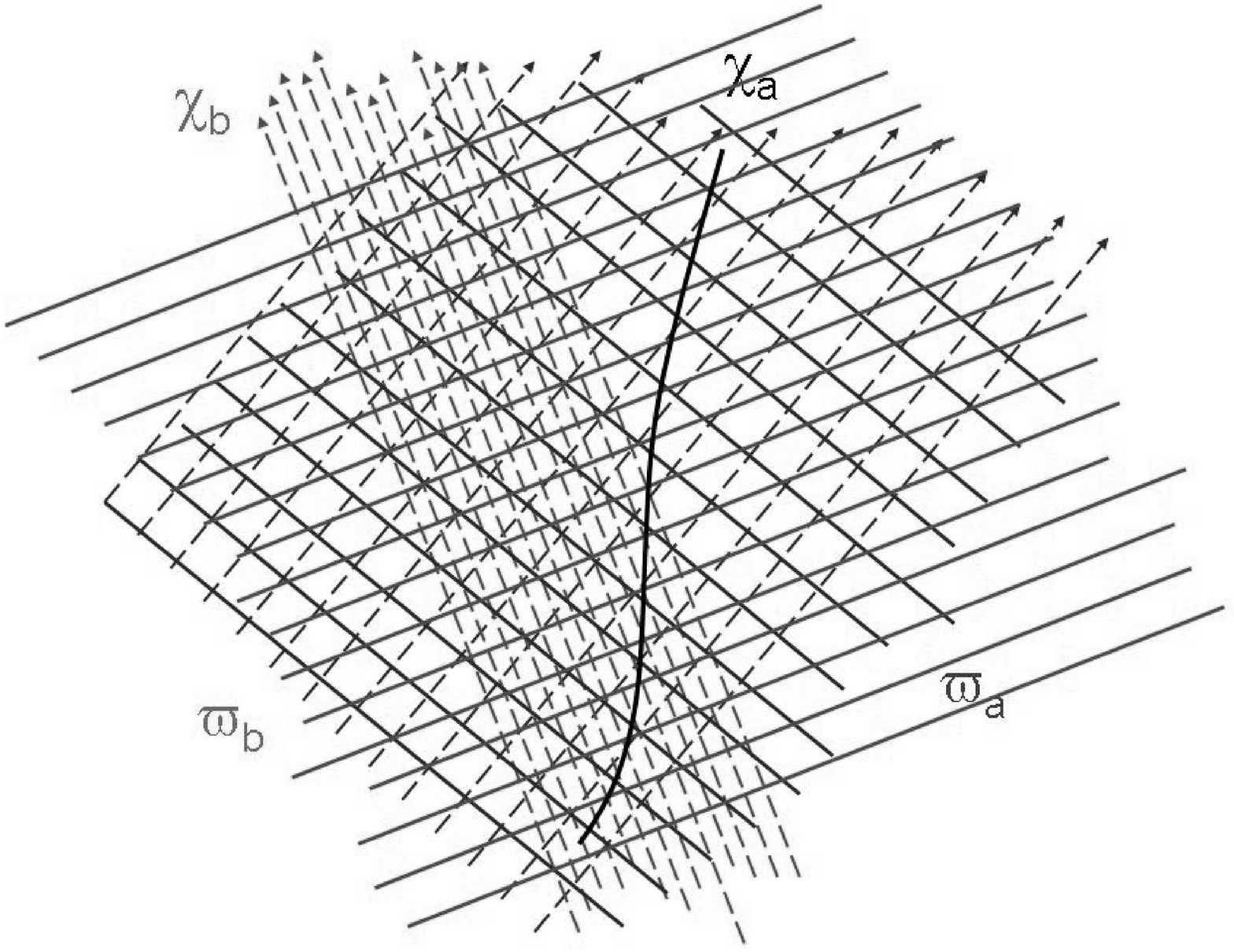}
\end{center}
\caption{An example of the grid mapping,
in this case, a flat 1+1 dimensional manifold. Vertically there is time,
horizontally space. Dashed lines are worldlines of electromagnetic signals;
the arrows show the directions of the null wavevectors $\protect\chi _{a}$
and $\protect\chi _{b}$. The symbols $\protect\varpi _{a,b}$ label the two
families of hyperplanes corresponding to the wavefronts. Each wavefront is
associated with an integer number; the knots in the resulting grid are
identified by a pair of integers: their null coordinates. The continuous
wavy line represents the world-line of an observer.}
\label{fig01}
\end{figure}

Suppose now to have a number ($\geq 4$) of sources of electromagnetic pulsed
signals located at infinity and at rest with respect to each other. Let us
assume also that our space-time is flat (Minkowski space-time). In these
conditions the four-dimensional trajectories of electromagnetic signals, as
well as of freely falling objects (inertial observers) are straight lines.
Each one of our sources is characterized by the frequency of its pulses and
by their direction in space. In a relativistic description this information
is contained in a four- vector, tangent to the space-time trajectory
(world-line) of the light rays. The four-vector of the $a$-th\footnote{%
Latin indices run from 1 to 4.} source may be written as
\begin{equation*}
\chi _{a}=\frac{2\pi }{T_{a}}(1,\hat{n}_{a})
\end{equation*}%
where $T_{a}$ is the period of the pulsar and $\hat{n}$ is a unit ordinary
vector in space (three-vector) whose components are the direction cosines in
any given reference frame. In relativity $\chi _{a}$ is the wave four-vector
of the electromagnetic signal from source $a$ and is a null vector because
it is self-orthogonal, i.e. the scalar product by itself is zero
\begin{equation}
\chi _{a}^{2}=\chi _{a}\cdot \chi _{a}=0  \label{due}
\end{equation}%
Under the general conditions we have chosen the wave vector from each source
is fixed (the same everywhere and at all times). If we wish to localize an
event with respect to an arbitrary origin, we may think to draw a radius
four-vector from the origin to the given event. In four, as well as in
three, dimensions a vector is characterized by its components with respect
to some basis. In our case we can choose as basis vectors four neither
collinear nor coplanar wave vectors from different sources. This will be a
typical null basis, so called because of the nature of its elements \cite%
{rovelli}. The components of the position four-vector with respect to this
basis will be called the null coordinates of the given event. We may
associate to each of the new basis vectors (which essentially marks a
direction in space-time) a family of conjugate three-dimensional
hypersurfaces proceeding more or less as we would do in three dimensions. In
this case we get a family of parallel null hyperplanes filling up the whole
space-time. In three dimensions one would obtain the wave fronts and the
concept may safely be extended to four dimensions. We then may single out
the set of hyperplanes that represent the wave fronts of the corresponding
pulses and label them by the ordinal numbers counted from an arbitrary
origin. Finally we cover the whole space-time with a grid made of the
fiducial hyperplanes. The situation is sketched in figure \ref{fig01}.

Considering now a receiver able to recognize and count the signals coming
from the various sources, the ordinal numbers used to label the subsequent
pulses will naturally be a set of null coordinates using the $\chi $ vectors
as their basis and having a single value over each specific null
hypersurface. Introducing the periods of each source, the position
four-vector of an event may then be written as
\begin{equation}
r=[(n+x)T]^{a}\chi _{a}  \label{tre}
\end{equation}%
The Einstein convention has been used\footnote{%
When two indices are \textquotedblleft contracted \textquotedblleft , i.e.
the same index appears once in upper position, once in lower position, a
summation has to be made over the whole range of the values of the index. In
our case from 1 to 4.}; $n$ is an integer and $x$ a fractional number,
usually ranging from 0 to 1, even though, as we shall see, it will be
allowed to assume values bigger than 1.

In practice, considering the dimensions and the definition of $\chi $, the
components of the position four-vector coincide with the phases of the waves
reaching the event from the various sources. The phase has the same value
over any given hypersurface.

While travelling through space-time (see fig.\ref{fig01}) an observer
\textquotedblleft crosses\textquotedblright\ successive hyperplanes and can
count the crossing events (arrival of the pulses). The start (the $0$) of
the reckoning will be arbitrarily chosen in correspondence of the arrival
of, say, a signal from source $a$, then the $0$ for source $b$ will be the
first $b$ pulse arriving after the $0$ of $a$, etc... If the receiver
carries its own clock, this can be used to measure the proper time span
between the arrival times from the various sources. In practice the receiver
will be able to reconstruct its own world-line with respect to the
infinitely far away sources and the arbitrarily chosen origin. We have
implicitly assumed that the frequencies of the pulses from the basis sources
are not too different from each other and the best choice would be to use
the smallest one as the reference (the $a$-th source). As said, the families
of hyperplanes build a space-time coordinated grid; the knots of the grid
are identified by quadruples of integer numbers.

A general event will of course not coincide with any of the knots of the
grid, but will be contained within a four-volume whose size, expressed in
metrical units, will be%
\begin{equation}
\Delta V=\frac{c^{2}}{4!}\varepsilon ^{abcd}\left( T^{2}\chi \right)
_{a}\wedge \left( T^{2}\chi \right) _{b}\wedge \left( T^{2}\chi \right)
_{c}\wedge \left( T^{2}\chi \right) _{d}  \label{quattro}
\end{equation}

The compact mathematical notation of eq. (\ref{quattro}) introduces the
speed of light $c$ and the completely anti-symmetric Levi-Civita tensor. In
practice (\ref{quattro}) is the determinant of a square 4$\times $4 matrix
whose rows are the vectors $(T^{2}\chi )_{a}$. Since the directions in space
of the basis null vectors will in general not be orthogonal to each other,
the size of the uncertainty volume depends also on the angles between the
positions of the sources in the sky. It is however possible to make a rough
estimate of the linear uncertainty in the position. In the case of
millisecond pulsars the typical edge of the uncertainty hypervolume, then
the localization uncertainty in space, would be very big, in the order of $%
\sim 10^{5}$ m. However, if the receiver is equipped with its own clock, as
we have assumed, the observer is able to measure the proper time span
between the arrival times of the signals, i. e. the interval (space-time
distance) between the intersections of his world-line and, say, the $n_{a}$%
-th hyperplane and the three successive $n_{b}$-th, $n_{c}$-th, $n_{d}$-th.
This information makes the reconstruction of the observer's world-line
possible with an accuracy determined by the precision of the onboard clock.
Using for instance a quartz oscillator one would have $\sim 10^{-9}$ s or 30
cm; using atomic clocks this could reduce to $\sim $ 3 cm or less.

\section{Detailed positioning within the grid}

In order to locate the arrival events at positions that do not coincide with
the knots of the grid, we may start from eq. (\ref{tre}). Suppose the
emission from the sources is continuous and the arrival times can be
determined with an arbitrary precision, then we may rewrite (\ref{tre}),
leaving aside the distinction in an integer and a fractional part,
\begin{equation}
r=\tau ^{a}\chi _{a}  \label{cinque}
\end{equation}%
A successive event along the world line of the receiver will be at%
\begin{equation}
r^{\prime }=\tau ^{\prime a}\chi _{a}=\tau ^{\prime 1}\chi _{1}+\tau
^{\prime 2}\chi _{2}+\tau ^{\prime 3}\chi _{3}+\tau ^{\prime 4}\chi _{4}
\label{sei}
\end{equation}

The flatness hypothesis allows us to write the displacement vector between
the two events as a simple difference vector%
\begin{equation}
\Delta r=r^{\prime }-r=\left( \tau ^{\prime 1}-\tau ^{1}\right) \chi
_{1}+\left( \tau ^{\prime 2}-\tau ^{2}\right) \chi _{2}+\left( \tau ^{\prime
3}-\tau ^{3}\right) \chi _{3}+\left( \tau ^{\prime 4}-\tau ^{4}\right) \chi
_{4}  \label{sette}
\end{equation}

If we suppose that the acceleration of the receiver is small enough to make
the change in velocity between $r$ and $r\prime $ negligible, i.e. that the
world-line is practically rectilinear between the two events, we have that
the norm of $\Delta r$ coincides with the corresponding proper time interval
$\Delta t$:%
\begin{equation}
\begin{array}{c}
\left( \Delta t\right) ^{2}=2g_{ab}\Delta t^{a}\Delta t^{b}+2g_{ac}\Delta
t^{a}\Delta t^{c}+2g_{ad}\Delta t^{a}\Delta t^{d} \\
+2g_{bc}\Delta t^{b}\Delta t^{c}+2g_{bd}\Delta t^{b}\Delta
t^{d}+2g_{cd}\Delta t^{c}\Delta t^{d}%
\end{array}
\label{otto}
\end{equation}

Eq. (\ref{otto}) is nothing else than the square of (\ref{sette}), i. e. the
scalar product by itself. In the formula we have introduced the fundamental
geometric operator which is needed in order to write the scalar products:
the metric tensor $g_{ab}$. Given a basis the elements of the metric tensor
in that representation are the scalar products between pairs of basis
vectors. In our case we have%
\begin{equation}
g_{ab}=\left( T\chi \right) _{a}\cdot \left( T\chi \right) _{b}  \label{nove}
\end{equation}

The $\chi $'s are null vectors, so all diagonal terms of the metric tensor
are zero:%
\begin{equation}
g_{aa}=T_{a}^{2}\chi _{a}\cdot \chi _{a}=0  \label{dieci}
\end{equation}%
Coming to a more realistic situation we remember that the signals consist in
a series of pulses, whose periods we have included into the basis vectors in
order to make them adimensioned. As we already did in eq. (\ref{tre}) we can
express the null coordinates of the event we are considering in terms of an
integer plus a fractional part, so that in general:%
\begin{equation}
\tau ^{a}=\left[ \left( n+x\right) T\right] ^{a}  \label{undici}
\end{equation}

Let us next consider a sequence of arrival times from respectively sources $%
a,b,c,d$ and label the corresponding events as 1,2,3,4,5,6,7,8. We choose
the first event as coinciding with the crossing of a hyper-surface belonging
to the $a$ family, so that $x_{a1}=0$ . The second event will be at the next
crossing of a $b$-hypersuface, the third will be at the encounter with the
subsequent $c$-hypersurface, and so on. The arrival times of the signals in
the sequence will correspond to the space-time positions:%
\begin{equation}
\begin{array}{c}
r_{1}=\left( n_{1}T\right) ^{a}\chi _{a}+\left[ \left( n_{1}+x_{1}\right) T%
\right] ^{b}\chi _{b}+\left[ \left( n_{1}+x_{1}\right) T\right] ^{c}\chi
_{c}+\left[ \left( n_{1}+x_{1}\right) T\right] ^{d}\chi _{d} \\
r_{2}=\left[ \left( n_{2}+x_{2}\right) T\right] ^{a}\chi _{a}+\left(
n_{2}T\right) ^{b}\chi _{b}+\left[ \left( n_{2}+x_{2}\right) T\right]
^{c}\chi _{c}+\left[ \left( n_{2}+x_{2}\right) T\right] ^{d}\chi _{d} \\
r_{3}=\left[ \left( n_{3}+x_{3}\right) T\right] ^{a}\chi _{a}+\left[ \left(
n_{3}+x_{3}\right) T\right] ^{b}\chi _{b}+\left( n_{3}T\right) ^{c}\chi _{c}+%
\left[ \left( n_{3}+x_{3}\right) T\right] ^{d}\chi _{d} \\
r_{4}=\left[ \left( n_{4}+x_{4}\right) T\right] ^{a}\chi _{a}+\left[ \left(
n_{4}+x_{4}\right) T\right] ^{b}\chi _{b}+\left[ \left( n_{4}+x_{4}\right) T%
\right] ^{c}\chi _{c}+\left( n_{4}T\right) ^{d}\chi _{d} \\
r_{5}=\left[ \left( n_{1}+1\right) T\right] ^{a}\chi _{a}+..... \\
r_{6}=....+\left[ \left( n_{1}+1\right) T\right] ^{b}\chi _{b}+..... \\
r_{7}=....+\left[ \left( n_{1}+1\right) T\right] ^{c}\chi _{c}+..... \\
r_{8}=....+\left[ \left( n_{1}+1\right) T\right] ^{d}\chi _{d}+.....%
\end{array}
\label{dodici}
\end{equation}

We may easily count the $n$'s but have no direct means to measure the $x$'s.
However if we suppose that the acceleration of the receiver is small enough
to allow for the identification of the world line, in a couple of periods of
the sources, with a straight line and if we carry on board the receiver a
clock, we are able to measure the proper time intervals between the $i$-th
and $j$-th arrival events, $t_{ij}$. With all this, trivial geometric
considerations lead to the values of our interest. It is%
\begin{equation}
\begin{array}{c}
x_{a1}=0,\text{ ~}x_{b1}=1-\frac{t_{12}}{t_{26}},\text{ ~}x_{c1}=1-\frac{%
t_{13}}{t_{37}},\text{ ~}x_{d1}=1-\frac{t_{14}}{t_{48}} \\
x_{a2}=\frac{t_{12}}{t_{15}},\text{ ~}x_{b2}=1,\text{ ~}x_{c1}=x_{c1}+\frac{%
t_{12}}{t_{37}},\text{ ~}x_{d2}=x_{d1}+\frac{t_{12}}{t_{48}} \\
x_{a3}=\frac{t_{13}}{t_{15}},\text{ ~}x_{b3}=1+\frac{t_{13}}{t_{26}},\text{ ~%
}x_{c3}=1,\text{ ~}x_{d3}=x_{d1}+\frac{t_{13}}{t_{48}} \\
x_{a4}=\frac{t_{14}}{t_{15}},\text{ ~}x_{b4}=1+\frac{t_{14}}{t_{26}},\text{ ~%
}x_{c4}=x_{c1}+\frac{t_{14}}{t_{37}},\text{ ~}x_{d4}=1%
\end{array}
\label{tredici}
\end{equation}
Moving the sequence forward by four steps, we can reconstruct the whole
world-line of the receiver.

\subsection{Accuracy}

Suppose the accuracy of the on board clock is $\pm \delta t$. It will be
reflected in an uncertainty on the values of the $x_{a}$'s. In the worst
case it would be:%
\begin{equation}
\left\vert \delta x_{a}\right\vert =4\left( \frac{1}{t_{i,i+4}}+\frac{%
t_{1,1+1}}{t_{1,1+4}^{2}}\right) \delta t  \label{quattordici}
\end{equation}

For example if we have $\delta t\sim 10^{-10}$ s\footnote{%
Reasonable for an atomic clock} and it is $t_{i,i+1}\approx t_{i,i+4}/4$
with $t_{i,i+4}\symbol{126}10^{-3}$ s\footnote{%
As it would be for millisecond pulsars}, the final uncertainty on the
coordinates would be
\begin{equation}
\left\vert \delta x_{a}\right\vert \approx 5\times 10^{-7}\text{ s}
\label{quindici}
\end{equation}%
Multiplying by the speed of light we get the equivalent uncertainty in the
positioning, which would be in the order of 150 m.

The situation becomes better if we allow for longer proper time intervals
(instead of $t_{i,i+4}$ put $t_{i,i+4n}$, being $n$ an integer). In turn the
maximum interval you may use in this approach is limited by the viability of
the linear world-line hypothesis, i.e. by the acceleration of the observer
with respect to the "fixed stars".

\section{Accelerated motion and perturbations}

According to the procedure described in the previous section the accuracy of
the final localization depends on the accuracy of the onboard clock. From (%
\ref{quattordici}) we have seen that the final result depends on the length
of the receiver's world-line over which a linear dependence on proper time
may be assumed. Let us then write the parametric equation of the world-line
from a given event up to second order in proper time. It is:%
\begin{equation}
\begin{array}{c}
x^{a}=u_{\left( 0\right) }^{a}\frac{t}{T_{a}}+\frac{1}{2}\frac{a^{a}}{T_{a}}%
t^{2} \\
x^{b}=.... \\
....%
\end{array}
\label{sedici}
\end{equation}

Here $u^{a}$ and $a^{a}$ are respectively the $a$-th components of the
four-velocity $u$ and four-acceleration $a$ of the observer at proper time $%
t=0$\footnote{%
The space components of the four-velocity are $u_{space}=\frac{v}{c\sqrt{%
1-v^{2}/c^{2}}}$, being $v$ the ordinary velocity of the observer; the time
component is obtained replacing $v$ in the numerator with 1. The
four-acceleration is obtained differentiating the four-velocity with respect
to the proper time.}.

The linearity hypothesis is acceptable as far as the second term in the
development is smaller than the clock precision. The condition is%
\begin{equation}
\frac{1}{2}\left\vert a^{a}\right\vert t^{2}\leq \left\vert u^{a}\right\vert
\delta t  \label{diciassettea}
\end{equation}

Eq. (\ref{diciassettea}) defines also the maximum time interval that can be
used in order to reduce the localization inaccuracy; for longer times the
deviation from linearity appears. It will be%
\begin{equation}
t_{\max }=\sqrt{2\left\vert \frac{u^{a}}{a^{a}}\right\vert \delta t}
\label{diciotto}
\end{equation}

Just to fix a rule of thumb estimate, suppose we have a receiver moving in a
flat space time with $a\sim 10$ m/s$^{2}$ three-dimensional acceleration and
$v\sim 3\times 10^{5}$ m/s velocity. If the direction cosines are with
respect to a Cartesian background reference frame, we have%
\begin{equation*}
\frac{u^{a}}{a^{a}}=\frac{c}{g}\frac{1-v\left( \cos \alpha \cos \alpha
_{a}+\cos \beta \cos \beta _{a}+\cos \gamma \cos \gamma _{a}\right) /c}{%
\mathcal{D}}
\end{equation*}%
The unlabeled angles refer to the direction cosines of the observer's
velocity, the others to the positions of the sources; $g$ is the value of
the three-dimensional acceleration. The quantity $\mathcal{D}$ stays for
\begin{equation}
\begin{array}{c}
\mathcal{D=}\frac{v/c}{1-v^{2}/c^{2}}\times \left( \cos \alpha \cos \alpha
_{g}+\cos \beta \cos \beta _{g}+\cos \gamma \cos \gamma _{g}\right)  \\
\times \left( 1-\cos \alpha \cos \alpha _{a}-\cos \beta \cos \beta _{a}-\cos
\gamma \cos \gamma _{a}\right)  \\
-\left( \cos \alpha _{a}\cos \alpha _{g}+\cos \beta _{a}\cos \beta _{g}+\cos
\gamma _{a}\cos \gamma _{g}\right)
\end{array}
\label{diciannove}
\end{equation}%
The $g$-labeled angles refer to the direction cosines of the
three-acceleration vector.

Just to make an estimate, let us introduce the numbers of our example. We
have%
\begin{equation*}
t_{\max }\approx 8\times 10^{-2}\text{ s}
\end{equation*}%
which corresponds to several periods of a millisecond pulsar, thus
validating the linearization process described in the previous section.

\subsection{A gravitational field}

Of course a deviation from the linearity of the observer's world-line can
also be due to the curvature of space-time i.e. to a gravitational field.
Let us exclude, for the moment, any gravitomagnetic effect\footnote{%
Gravitomagnetic effects are due to the rotation of the source of gravity and
are extremely small.}, and consider an approximated line element\footnote{%
Space-time distance between two arbitrarily near events} like%
\begin{equation}
ds^{2}=g_{00}d\tau ^{2}-g_{ss}\left( dX^{2}+dY^{2}+dZ^{2}\right)
\label{venti}
\end{equation}%
represented in a Cartesian reference frame with space coordinates $X$,$Y$,$Z$%
.

In such a space the modulus of the space part of a null four-vector, is
given by
\begin{equation}
\left\vert k_{a}\right\vert =\frac{2\pi }{T_{a}}c\sqrt{\left\vert \frac{%
g_{00}}{g_{ss}}\right\vert }  \label{ventuno}
\end{equation}%
The $g_{00}$ and $g_{ss}$ factors account for the presence of the
gravitational field.

In a weak field, which is the usual condition almost everywhere in the
universe, we may assume%
\begin{equation}
\begin{array}{c}
g_{00}=1+\Phi \\
g_{ss}=-\frac{1}{g_{00}}\simeq -1+\Phi%
\end{array}
\label{ventidue}
\end{equation}%
being $\Phi <<1$ the (static) gravitational potential divided by $c$. We
have then%
\begin{equation}
\left\vert k_{a}\right\vert =\frac{2\pi }{T_{a}}c\left( 1+\Phi \right)
\label{ventitre}
\end{equation}

Once we have introduced these notations and this approximation it is
possible to convert the metric tensor to its form with respect to the null
basis. Then the metric tensor can be used to compute scalar products up to
first order in the perturbation represented by $\Phi $.

Remember that, by definition, the velocity four-vector is a unit vector and
let us start considering the identity:%
\begin{equation}
1=g_{ab}u^{a}u^{b}  \label{ventiquattro}
\end{equation}%
Let us then perturb it with respect to the flat case in the absence of
non-gravitational accelerations. We have%
\begin{equation}
0=u^{a}u^{b}\delta g_{ab}+2g_{ab}u^{a}\delta u^{b}  \label{venticinque}
\end{equation}%
To first order in $\Phi $ this becomes%
\begin{equation*}
0=\Phi g_{ab\left( 0\right) }u_{\left( 0\right) }^{a}u_{\left( 0\right)
}^{b}+2g_{ab\left( 0\right) }u_{\left( 0\right) }^{a}\delta u^{b}
\end{equation*}%
i.e.%
\begin{equation}
0=\Phi +2g_{ab\left( 0\right) }u_{\left( 0\right) }^{a}\delta u^{b}
\label{ventiosei}
\end{equation}%
The label $(0)$ refers to the flat space-time (no gravitational field)
values and use has been made of (\ref{ventiquattro}).

Solving eq. (\ref{ventiosei}) we get%
\begin{equation}
\delta u^{b}=-\frac{\Phi }{2}u_{\left( 0\right) }^{b}  \label{ventiosette}
\end{equation}%
We have used again the identity (\ref{ventiquattro}).

Let us consider now two successive events separated by the proper time
interval $dt$ and let us call $q^{a}$ the $a$ component of the purely
gravitational acceleration. We can write:%
\begin{equation}
\begin{array}{c}
u^{a\prime }-u^{a}=du^{a}=q^{a}dt=d\left( \delta u^{a}\right) \\
=-\frac{u_{\left( 0\right) }^{a}}{2}d\Phi =-\frac{u_{\left( 0\right) }^{a}}{2%
}\left( Tdx\right) ^{b}\nabla _{b}\Phi%
\end{array}
\label{ventotto}
\end{equation}%
From (\ref{ventotto}) we see that%
\begin{equation}
q^{a}=-\frac{u_{\left( 0\right) }^{a}}{2}u_{\left( 0\right) }\cdot
\triangledown \Phi  \label{ventinove}
\end{equation}%
The appropriated variables for the calculation of the gradient are the $%
(xT)^{a}$.

Introducing the uncertainty $\delta t$ in the proper time measurement and
using the second order development of the world-line (now a geodetic line) (%
\ref{sedici}) we see that the gravitational effect emerges out of the
experimental uncertainty when:%
\begin{equation}
u_{\left( 0\right) }^{a}\delta t\leq \frac{1}{4}u_{\left( 0\right)
}^{a}\left\vert u_{\left( 0\right) }\cdot \nabla \Phi \right\vert t^{2}
\label{trenta}
\end{equation}%
i.e.
\begin{equation}
\left\vert u_{\left( 0\right) }\cdot \nabla \Phi \right\vert \geq 4\frac{%
\delta t}{t^{2}}  \label{trentuno}
\end{equation}

\subsection{Distance and movement of the source}

So far we have assumed "fixed" sources located at infinity. Let us first
examine what can the effect of a finite distance be in the case of pulsars.
Geometrically the equal phase ordinary surfaces are no more plane and over a
distance $l$ the difference in position is $\sim l%
{{}^2}%
/R$ being $R$ the distance of the source from the receiver. Let us consider
the order of magnitude of a typical pulsar's distance $R\sim 10^{3}$ light
years $\sim 10^{19}$ m: we see that the induced position error (when
neglecting the curvature and assuming the distance to be infinite) stays
below $\sim 1$ cm for distances up to $l\sim 10^{8}$ m, comparable with the
Earth-Moon distance.

As for the proper motion of the pulsars, one with respect to the other, it
affects the basis null vectors producing an additional proper time
dependence and simulating a contributed "acceleration". We may express it in
a Cartesian reference frame, writing:%
\begin{equation}
a_{p}^{a}=\left. \frac{\partial u^{a}}{\partial t}\right\vert _{\text{proper
motion}}=-\overset{i=4}{\underset{i=1}{\dsum }}g_{ii}u^{i}\tan \alpha \frac{%
\partial \alpha }{\partial t}  \label{trentadue}
\end{equation}%
The $\alpha $ is any of the angular coordinates of the source in the sky.

A similar contribution comes from the slow decay of the pulsar's frequency.
Now it is%
\begin{equation}
a_{T}=\left. \frac{\partial u^{a}}{\partial t}\right\vert _{\text{period}}=%
\frac{u_{\left( 0\right) }^{a}}{T^{a}}\frac{\partial T^{a}}{\partial t}
\label{trentatre}
\end{equation}

In any case we may remark that the decay law of the period as well as the
proper motion in the sky are, for most pulsars, very well known, so that the
effects represented by eq.s (\ref{trentadue}) and (\ref{trentatre}) can be
treated as systematic and are easily accounted for.

\section{ Conclusion}

Despite the apparent complexity of the mathematics we have displayed, here
we have developed a very simple method for the use of pulsars for
localization purposes. In fact the local measurements to be done are just
time measurements that can be obtained with great accuracy, and the basic
formulae (\ref{dodici}) and (\ref{tredici}) are plain linear relations. The
effects of acceleration, either from gravitational or non-gravitational
origin, appear at the second order and again calculations are very simple.
The final accuracy in the positioning would easily be in the meter range,
being essentially controlled by the precision of the clock used by the
observer.

Of course the main problem with pulsars, which is the extreme weakness of
their signals, stays there, and requires appropriate technological
improvements in the detection devices. I would also remark that we thought
of millisecond pulsars, however the method I have sketched is in no way
depending on the class of pulsars we choose. Everything works equally well,
for instance, with X-ray pulsars or even with second pulsars.

Finally, in principle the same approach I have presented can be applied to
sources other than pulsars. If for instance we consider satellites orbiting
the Earth, much like in the GPS, the method still works, provided you allow
for the time dependence of the direction cosines of the null four-vectors of
the null basis. Now these cosines would depend on proper time according to
the space-time orbit of the satellites, which we may think to accurately
know.

The method is fully relativistic and it automatically accounts for
gravitational redshift and speed effects. Furthermore there is no need for
synchronization between the signals from the different sources, nor have
they to have the same frequency.

Summing up, I think this approach would be the most modern and appropriate,
after one century of relativity, performing a new technological and
methodological Copernican revolution, transferring the basic frame from the
Earth to the very space-time with its intrinsic properties and curvature.

\section{Acknowledgments and credits}

Various members of the RELGRAV group of the Physics Department of the
Politecnico di Torino collaborated to the present work which is part of a
project for the development of a fully relativistic positioning system. They
are Emiliano Capolongo, Roberto Molinaro and Matteo Luca Ruggiero.

Our research has been supported by Piemonte local government within the
MAESS-2006 project "Development of a standardized modular platform for
low-cost nano- and micro-satellites and applications to low-cost space
missions and to Galileo" and by ASI.

\end{document}